\documentclass[12pt,a4paper]{article}
\usepackage{latexsym,amsmath,amssymb}
\date{}
\numberwithin{equation}{section}
\def\FF{{\mathcal{F}}}
\begin{document}
\title{Quantum field components of supersymmetric multiplets}
\author{Florin Constantinescu\\ Fachbereich Mathematik \\ Johann Wolfgang Goethe-Universit\"at Frankfurt\\ Robert-Mayer-Strasse 10\\ D 60054
Frankfurt am Main, Germany
\and{} 
G\"unter Scharf\\ Institut f\"ur Theoretische Physik, Universit\"at Z\"urich \\
Winterthurerstr. 190 \\
CH 8057 Z\"urich, Switzerland}
\maketitle
\begin{abstract}

We adress the problem of Fock space representations of (free) multiplet component fiels encountered in supersymmetric quantum field theory insisting on positivity and causality. We look in detail on the scalar and Majorana components of the chiral supersymmetric multiplet. Several Fock space representations are introduced. The last section contains a short application to the supersymmetric Epstein-Glaser method. The present paper is written in the vane of axiomatic quantum field theory with applications to the causal approach to supersymmetry.
\end{abstract}

\section{Introduction}

In this paper we consider the representation of the $N=1$ supersymmetric (SUSY) algebra in four dimensions \cite{WB,W}
\begin{gather} 
\{Q_a,\bar{Q}_{\bar{b}}\} = 
2{\sigma^{\mu}}_{a \bar{b}}P_{\mu}   \notag \\ 
[ Q_a , Q_b ] = [ \bar{Q}_{\bar{a}} , \bar{Q}_{\bar{b}} ] =0  \\ 
[ Q_a , P_{\mu} ] = [ \bar{Q}_{\bar{a}} , P_{\mu} ]=0  \notag \\
[P_{\mu},P_{\nu}]=0 \notag
\end{gather} 
on quantum (scalar) chiral superfields. 
Generally supersymmetry algebras can be represented on superfields which beside other properties satisfy causality (see for instance \cite{CSp}) generating in this way (super) field supersymmetry algebras. We will make difference between these two types of algebras like for instance the SUSY algebra (1.1) with generators $Q_a,  \bar{Q}_{\bar{b}} $ and the causal (free) superfield algebras with superfield generators $\Phi(x,\theta, \bar{\theta} ) $ depending on the (super-)variables $x,\theta,\bar{\theta}$ which can be conveniently written in terms of multiplet components. In the first part of this paper we restrict ourself to this older component approach disregarding for the moment the anticommuting variables. In this approach a (scalar) chiral superfield requires for its description two (complex) scalar and a Majorana field which are usually called the scalar and the Majorana components of the multiplet respectively. We study (without too much reference to supersymmetry) both scalar and Majorana components regarding them separately as causal algebras of operator valued distributions and giving their explicit Fock space representations. Let us consider the scalar component of the chiral supersymmetric multiplet consisting of two coupled scalar fields $A(x)$ and $F(x)$ \cite{WB}. A priori $A(x)$ and $F(x)$ are not independent such that the scalar component consisting of $A(x)$ and $F(x)$ cannot be realized through tensor product. In Section 3 we give a rigorous definition of the scalar component as a causal free field algebra which we subsequently represent in a properly constructed Fock space. If the equations of motion are satisfied this algebra degenerates to the usual (complex) scalar field with $F(x)=-mA^*(x) $. Similarly in section 4 we construct the Majorana causal  field algebra together with its explicit Fock space representation. Section 5 contains a short application to the non-renormalization theorem in the causal approach. In order to give an idea of the procedure used in this paper we include the preparatory section 2 which might be of some independent interest. \\

The aim of this paper is twofold: first it shows how to put the older component approach to supersymmetry on the firm basis of rigorous (=``axiomatic") free quantum field theory and second prepares the ground for the causal perturbation approach \cite{EG,S,Sg} to superymmetry for some simple as well as more involved models. We insist on positivity and causality of our Fock space representations. Realizing positivity and causality in explicit Fock space representations (i.e. exhibiting Fock space representations for causal field algebras appearing in the SUSY context) is the marking point of our study.\\

Before starting let us remark that the pedagogical example in the  first part of  Section 2 refers to canonical commutation relations. It was introduced there to simply illustrate the appearence of (harmless) zero-vectors for the Fock space representations of the multiplet components. They can be easily disregarded by factorization. Although possible we didn't try to follow the route of canonical quantization in order to construct our multiplet components; we prefer to define the multiplet components as Fock space representations of certain field algebras of causal commutation relations. This is an input for developping SUSY quantum field theory in the causal approach (without reference to canonical quantization or to the functional integral approach) as shown in section 5. In the more traditional way of looking at the problem the reader will easily find out that our approach is ``off shell". In particular the equations of motions (i.e. the field equations) are not a priori imposed on the operator valued distributional quantum multiplet components. Closing or not closing causal field superalgebras through field equations (i.e. working ``on" or ``off shell") is important when formulating models of rigid as well as extended supersymmetry cf. \cite{WB} p.22 and \cite{West} p.21.

\section{A pedagogical example}

In the first part of this section we start with a simple example by considering the following algebra of commutation relations ( $p,q \in  \mathbb R^d $):

\begin{gather}
[a(p), a^*(q)]=[a(p), b^*(q)]=[b(p),b^*(q)]=[b(p),a^*(q)]=\delta (p-q)  \notag \\
[a(p),a(q)]=[a(p),b(q)]=[b(p),b(q)]=0   \\
[a^*(p),a^*(q)]=[a^*(p),b^*(q)]=[b^*(p),b^*(q)]=0 \notag
\end{gather}
Remark that if $b(p)=a(p)$ then (2.1) reduces to the usual (Heisenberg) commutation relations. We want to construct the Fock space representation of (2.1). In order to do this consider the complex linear space $E$ of $L^2(\mathbb R^d )$-vector functions 
$f(p)= 
\begin{pmatrix} f_1(p)\\
                f_2(p)
\end{pmatrix}$ 
with the sesquilinear form 
\begin{equation}
(f,g)=\int (\bar f_1(p)+\bar f_2(p))(g_1(p)+g_2(p))dp
\end{equation}
which is non-negative:
\[
(f,f)=\int \left |f_1(p)+f_2(p) \right|^2 dp \geq 0
\]
being zero if $ f_1+f_2=0$ almost everywhere in $ \mathbb R^d $.
Certainly this space can be turned into a Hilbert space by the usual factorization procedure but we prefer to work with pre-Hilbert spaces (the zero vectors turn out to be harmless and the representation to follow factorizes). \\

Now put $\mathcal F^1=E$, consider the symmetric tensor product $\mathcal F^n=\mathcal F^1 \otimes \cdots \otimes \mathcal F^1 $ and write
$ \mathcal F= \oplus_ n \mathcal F^n  $ with its natural sesquilinear form and vacuum $\Omega \in \mathcal F^0=\mathbb{C} $. We describe elements in $\mathcal F$ up to symmetrization. Elements in $\mathcal F^n $ are tensor product functions

\begin{gather} 
\psi=(\psi^n) \equiv (\psi_{\mu_1,\cdots,\mu_n}^n(p_1,\cdots,p_n)), \\
\phi=(\phi^n) \equiv (\phi_{\nu_1,\cdots,\nu_n}^n(p_1,\cdots,p_n)) 
\end{gather}
with $\mu_1,\cdots,\mu_n=1,2 ;\quad \nu_1,\cdots,\nu_n=1,2$. The simple tensors in $\mathcal F^n$ are of the form
$f^1\otimes\cdots\otimes f^n=(f_{\mu_1}^1(p_1)\cdots f_{\mu_n}^n(p_n))$
with $\mu_1,\ldots,\mu_n=1,2$. \\

From now on, if clear from the context, we leave out the indices $\mu,\nu$ writing for instance instead of (2.3),(2.4) simply

\begin{equation} 
\psi^n=(\psi^n(p_1,\cdots,p_n)),\quad
\phi^n=(\phi^n(p_1,\cdots,p_n)) 
\end{equation}
or even
\[
\psi^n=\psi^n(p_1,\cdots,p_n),\quad
\phi^n=\phi^n(p_1,\cdots,p_n)
\]
For example the induced sesquilinear form in $\FF^n $  will be

\begin{equation}
(\psi^n,\phi^n)= \int\cdots\int(\sum_{\mu_i}\bar\psi_{\mu_1,\cdots,\mu_n}^n(p_1,\cdots,p_n))
(\sum_{\nu_i}\phi_{\nu_1,\cdots,\nu_n}^n(p_1,\ldots,p_n))dp_1\ldots dp_n
\end{equation}
and hence for simple tensors

\begin{equation}
(f^1\otimes\cdots\otimes f^n,g^1\otimes\cdots\otimes g^n)= \int\cdots\int\prod_{i=1}^n( \sum_{\mu_i} \bar f_{\mu_i}^i(p_i))\prod_{i=1}^n( \sum_{\nu_i}g_{\nu_i}^i(p_i))dp_1\ldots dp_n
\end{equation}
The graduation in $\FF$ implies $ (\phi^n,\phi^m)=0 $ for $ n\neq m $. \\

We want to represent the algebra (2.1) on $\FF$. For this we introduce some notations. First we leave out the upper index for elements in $\FF^n $. Second, by abuse, we make no difference between sesquilinear form and scalar product. Third there is an alternative way to write down the scalar product. For instance 
\[
(f(p),\psi(p,p_1,\ldots,p_n))=\int \bar f(p) \, \delta \, \psi(p,p_1,\ldots,p_n))dp
\]
Here $\delta$ is the matrix
\[ \delta =
\begin{pmatrix}
1 & 1 \\
1 & 1
\end{pmatrix}
\]
and it is understood that the contraction is over the first index in $\psi $ (and f). For instance if $\psi(p)=g(p) \in \FF^1 $ then
\begin{equation}
(f(p),g(p))= \int \bar f(p)\, \delta \, g(p)dp=\int( \bar f_1(p)+ \bar f_2(p))(g_1(p)+g_2(p))dp
\end{equation}
Remark that the scalar product not only reduces the number of variables but consistently reduces the rank of the tensor too. This is precisely what we want by using the scalar product as ``annihilation" part of our representation (the tensor product itself which at the same time raises the number of variables and the rank of the tensor will produce the ``creation" part).\\

On $ \FF $ we define the following operators for 
$f=\begin{pmatrix}f_1 \\f_2 \end{pmatrix} ,f_i \in L^2(\mathbb R^d),i=1,2 $.\\

\begin{align}
(\alpha(f)\psi)(p_1,\ldots,p_n) &=(n+1)^{\frac{1}{2}}(f(p),      \psi(p,p_1,\ldots,p_n) \\
(\alpha^*(f)\psi)(p_1,\ldots,p_n) &=n^{-\frac{1}{2}} \sum_{i=1}^n f(p_i)\psi(p_1,\ldots,\hat p_i,\ldots,p_n),
\end{align}
where in the first relation the scalar product and in the second relation the tensor product of $f$ with $\psi$ appears. They satisfy besides $\alpha(f)^*=\alpha^*(\bar f)$ the commutation relations:

\begin{align}
[\alpha(f), \alpha(g)] &=[\alpha^*(f),\alpha^*(g)]=0 \\
[\alpha(f),\alpha^*(g)] &=(\bar f,g)
\end{align}
where certainly $(f,g)= \int \bar f(p) \, \delta \, g(p)dp $. \\

For $f \in L^2(\mathbb R^d)$ define the operators
\begin{eqnarray}
a(f) = \alpha \begin{pmatrix} f \\0 \end{pmatrix} ,\quad b(f) = \alpha \begin{pmatrix} 0 \\f \end{pmatrix} \\
a^*(f) = \alpha^* \begin{pmatrix} f \\ 0 \end{pmatrix} ,\quad b^*(f) = \alpha^* \begin{pmatrix} 0\\f \end{pmatrix}
\end{eqnarray} 
From linearity in (2.13),(2.14) we have $\alpha(f)=a(f_1)+b(f_2)$ for $f=\begin{pmatrix} f_1 \\f_2 \end{pmatrix} $.
It is easy to see that (2.9) together with (2.10) produce a (smeared out) representation of the algebra (2.1). In particular the relation $[a(f),b^*(g)]=[\alpha \begin{pmatrix} f\\0 \end{pmatrix},\alpha ^* \begin{pmatrix}0\\g\end{pmatrix}]=(\bar f,g)$ follows from the fact that $\delta $ is non-diagonal producing a coupling of $\begin{pmatrix}f\\0\end{pmatrix}$ and $\begin{pmatrix}0\\g\end{pmatrix}$ and consequently the coupling of $ a $ and $ b $ in (2.1).
On the vacuum $ \Omega \in \FF  $ we have for $ f=\begin{pmatrix}f_1\\f_2 \end{pmatrix} $:

\begin{equation}
\alpha(f)\Omega=0 
\end{equation}
 such that, as expected for $ f \in L^2(\mathbb R^d)$
\begin{equation}
a(f)\Omega =b(f)\Omega=0  
\end{equation}

So far it seems that there is nothing special with this representation. Though it has a marking point: the Fock space $\FF$ in which we represent has zero-vectors which were pointed out above. It is not difficult to see that the set of zero vectors is left invariant by the algebra (2.1) such that we can eliminate them by the standard procedure obtaining a bona-fide Fock space representation of the algebra (2.1).\\
Now  for $d=4$ we can use $a^\#(p),b^\#(p), \, p=(p_0,\bar p),\bar p \in \mathbb R^3 $ in order to construct fields 

\begin{eqnarray}
A(x)=\int\frac{d\bar p}{(2\pi)^\frac{3}{2}(2p_0)^\frac{1}{2}}[a^*(p)e^{ipx}+a(p)e^{-ipx}]\\
F(x)=\int\frac{d\bar p}{(2\pi)^\frac{3}{2}(2p_0)^\frac{1}{2}}[b^*(p)e^{ipx}+b(p)e^{-ipx}]
\end{eqnarray}
where $ x \in \mathbb R^4  $ and $ px=p_0x_0-\bar p\bar x,p_0=\sqrt{p^2+m^2}$. Elementary considerations show that $A(x)$ and $F(x)$ satisfy the following (causal) field algebra

\begin{equation}
\begin{split}
&[A(x),A(y)]=[A(x),F(y)]=[F(x),F(y)]= \\
&=\frac{1}{2(\pi)^3}\int dp \, \delta  \, (p^2-m^2)\text{sign}(p^0)e^{-ip(x-y)}= \\
&=-iD(x-y)
\end{split}
\end{equation}
where the Pauli-Jordan function (distribution) is given by $D(x)=D^+(x)+D^-(x)=D^+(x)-D^+(-x);D^-(x)=-D^+(-x) $ where

\begin{gather}
D^+(x)=\frac{i}{(2\pi)^3}\int dp\theta(p^0)\delta(p^2-m^2)e^{-ipx} ,\\
D^-(x)=\frac{-i}{(2\pi)^3}\int dp\theta(-p^0)\delta(p^2-m^2)e^{-ipx} \\
D(x)=\frac{i}{(2\pi)^3}\int dp\text{sign}(p^0)\delta(p^2-m^2)e^{-ipx}
\end{gather}
Although both $-iD^+(x)$ and $iD^-(x)$ are positive definite their difference $-iD(x)$ is not. Instead it is causal, a property which in turn is not shared by $-iD^{\pm}(x)$.
The interesting point in (2.19) is the commutator $[A(x),F(y)]=-iD(x-y)$ showing that the scalar fields A(x) and F(y) are in fact not independent. \\

We would like to represent the pair $ A(x),F(x)$ in a Fock space constructed over functions on $\mathbb R^4$. This is certainly possible by the procedure of canonical quantisation starting with (2.17),(2.18) and the Fock construction we have already performed but at this moment we want to change attitude. This is an important point which we want to motivate. The couple $ A(x),F(x)$ (2.19) does not appear in physical applications. In the next sections and in forthcoming work we will study or comment on some other component fields which really appear as parts of supersymmetric multiplets. They might be of increasing complexity and a construction of them by canonical quantization seems to be obscure. On the other side we can find out the (causal) field algebra of the multiplet in question by just algebraic consideration starting with the supersymmetry and some pivotal component field which is gradually inflated into the whole algebra. This suggests looking for Fock space representations of the (causal) multiplet algebra itself starting from scratch. We will do this for our example (2.19) proceeding to more interesting cases in the next sections. \\

Before starting we still have some comments on the notations. We write

\begin{gather*}
\Delta^{\pm}(p)=\theta(\pm p^0)\delta(p^2-m^2) \\
\Delta^-(p)=\Delta^+(-p) \\
\text{sign}(p^0)\delta(p^2-m^2)=\Delta(p)=\Delta^+(p)-\Delta^-(p)=\Delta^+(p)-\Delta^+(-p) \\ 
\Delta(-p)=-\Delta(p)\\
\underline\Delta^+(p)=\delta \Delta^+(p) ,\quad \delta= \begin{pmatrix}1&1\\1&1\end{pmatrix}
\end{gather*}
The functions $D^\pm(x),D(x)$ and $\Delta^\pm(p),\Delta(p)$ are connected by the Fourier transform
\begin{equation}
\tilde{f}(p)=\frac{1}{(2\pi)^2}\int f(x)e^{ipx}dx; \quad px=p_0x_0-\bar p \bar x \qquad x,p \in \mathbb R^4
\end{equation} 
Indeed

\begin{gather}
D^+(x)=\frac{i}{(2\pi)^3 }\int e^{-ipx}\Delta^+(p)dp \\
D^-(x)=\frac{-i}{(2\pi)^3} \int e^{-ipx}\Delta^-(p)dp \\
D(x)=D^+(x)+D^-(x)=\frac{i}{(2\pi)^3}\int e^{-ipx}\Delta(p)dp
\end{gather}
We also have for further use

\begin{gather}
\bar D^+(x)=\frac{-i}{(2\pi)^3}\int e^{ipx}\Delta^+(p)dp= \\
=\frac{-i}{(2\pi)^3}\int e^{-ipx}\Delta^+(-p)dp=
\frac{-i}{(2\pi)^3}\int e^{-ipx}\Delta^-(p)dp=D^-(x)\\
D(-x)=-D(x) \qquad (\partial_{\mu} D)(-x)=\partial_{\mu} D(x)
\end{gather}

Although the algebra (2.19) is considered in the x-space we will describe our Fock space over functions in the 4-dimensional p-space. They should be Fourier transforms $\tilde f(p)$ of the form (2.23). In order to simplify notations we take permission to leave out the tilde and write by abuse $f(p)$ instead of $\tilde f(p) $. If clear from the context we also write $\Delta^+(p)$ instead of $ \underline \Delta^+(p) $ (as for instance in the next equation).\\

Consider the linear space $E=S(\mathbb R^4) \otimes\mathbb C^2 $ of Schwartz vector functions $\begin{pmatrix}  f_1 \\f_2 \end{pmatrix},  f_i \in S(\mathbb R^4),  i=1,2$ with the sesquilinear form

\begin{gather}
(f,g)=(f,\Delta^+g)=\sum_{\mu ,\nu}^{2} \int \bar f_\mu (p) \Delta_{\mu\nu}^+(p)g_\nu(p)dp \notag \\
=\int(\sum_\mu \bar f_\mu (p))(\sum_\nu g_\nu(p))\Delta^+(p)dp
\end{gather}
We use the linear space $E$ as usual in order to construct our (symmetric) Fock space as above. The next result is the following Fock space representation of the algebra (2.19) where some factors of $(2\pi)^3$ in passing from $D$ to $\Delta $ were simply left out. It extends the Fock space representation of the scalar neutral field \cite{J}:

\begin{gather}
(A(f)\psi )(p_1,\ldots,p_n)=n^{-\frac{1}{2}}\sum_{i=1}^{n}\begin{pmatrix}f(p_i)\\0\end{pmatrix}\psi(p_1, \ldots ,\hat{p}_i, \ldots ,p_n) \notag \\ +(n+1)^{\frac{1}{2}}
(\begin{pmatrix}\bar{f}(-p)\\0\end{pmatrix},\psi (p,p_1,\ldots ,p_n))\\
(F(f)\psi )(p_1, \ldots ,p_n)=n^{-\frac{1}{2}}\sum_{i=1}^{n}\begin{pmatrix}0 \\ f(p_i)\end{pmatrix}\psi(p_1,\ldots,\hat p_i,\ldots,p_n) \notag \\ +(n+1)^\frac{1}{2}(\begin{pmatrix}0 \\\bar f(-p)\end{pmatrix},\psi(p,p_1,\ldots,p_n)
\end{gather}
where following our convention $\bar f(-p)$ means $ \bar{\tilde f}(-p)=\tilde{\bar f}(p)$. \\
Remark how the coupling between $A$ and $F$ is realized through the tensor and scalar product in (2.31) and (2.32). Again there is an invariant set of zero vectors. By standard factorization procedure we obtain the bona-fide Fock representation of causal algebra (2.19). It is now time to pass to more interesting causal algebras encountered in the frame of supersymmetric multiplets.

\section{The scalar component fields of the chiral multiplet}

The scalar component fields of the chiral multiplet provide the simplest example of component fields appearing in $N=1$ SUSY. It is defined through the algebra \cite{WB}  :
\begin{gather}
[A(x),A^*(y)]=-iD(x-y) \\
[F(x),F^*(y)]=im^2D(x-y) \\
[A(x),F(y)]=[F^*(x),A^*(y)]=imD(x-y) \\
[A(x),A(y)]=[F(x),F(y)]=[A(x),F^*(y)]=[A^*(x),F(y)]=0
\end{gather}
For mathematical convenience we take $m=1$. The algebra (3.1)-(3.4) describes two coupled scalar fields. The identification $F=-mA^*$ which is algebraically consistent with (3.1)-(3.4) is not assumed. In supersymmetry it is a consequence of the equations of motion which we do not assume a priori (cf. the similar situation for Majorana component in the next section).\\

In order to find a Fock space representation of the above system of two coupled complex scalar fields we comment first on the Fock space representation of one complex scalar field $\Phi(x)$. Because the complex scalar field describes particles of different charges $e=\pm 1 $ the simplest idea is to work with two-component wave functions $\psi(x,e)$ which depend on x and $e=\pm 1$ going through tensor products $\FF \otimes \FF^c $ required by the Fock construction \cite{SW},\cite{BLT}. Here c stays for antiparticle. Another idea for the case under consideration is to decompose the Fock space into 

\begin{gather}
\FF \otimes\FF^c=\oplus_{n,m=1}^{\infty} \FF^n\otimes \FF^{cm},  \FF=\oplus_n\FF^n, \FF^c=\oplus_m\FF^{cm}  
\end{gather}
The spaces $\FF^{n,m}=\FF^n\oplus\FF^{cm} $ consist of functions $f(p_1,\ldots, p_n;q_1,\ldots, q_m) $ symmetric in $p_1,\ldots,p_n$ and $q_1,\ldots,q_m$ separately but not necessarily fully symmetric. For the convenience of the reader we write down the representation of the complex scalar field $\Phi(x)$ subjected to the causal field algebra 

\begin{gather}
[\Phi(x),\Phi(y)]=[\Phi^*(x),\Phi^*(y)]=0, \\
[\Phi(x),\Phi^*(y)]=-iD(x-y)  
\end{gather}

It is
\begin{gather}
(\Phi(f)\psi)(p_1,\ldots,p_n;q_1,\ldots,q_n)= \\ \notag
= n^{-\frac{1}{2}}\sum_{i=1}^nf(p_i)\psi(p_1,\ldots,\hat p_i,\ldots,p_n;q_1,\ldots,q_m)+ \\ \notag
+(m+1)^\frac{1}{2}(\bar f(-q),\psi(p_1,\ldots,p_n;q,q_1,\ldots,q_m)) \\
(\Phi^*(f)\psi)(p_1,\ldots,p_n;q_1,\ldots,q_n)= \\ \notag
= m^{-\frac{1}{2}}\sum_{i=1}^mf(q_i)\psi(p_1,\ldots,p_n;q_1,\ldots,\hat q_i,\ldots,q_m)+ \\ \notag
+(n+1)^\frac{1}{2}(\bar f(-p),\psi(p,p_1,\ldots,p_n;q_1,\ldots,q_m))
\end{gather}
Here the Fock space is constructed over Schwartz space functions $f(p)$ and the scalar product is
\begin{equation}
(f(p),g(p)=\int \bar f(p)\Delta^+(p)g(p)dp
\end{equation}
(a disturbing $(2\pi)^{-3}$ factor in passing from $\Delta(p) $ to $D(x)$ was simply left out). One can verify that in the proper constructed Hilbert space one has ${\Phi(f)}^*=\Phi^*(\bar f)$ where $\bar f$ means complex conjugation and $ \Phi(f)^* $ is the operator adjoint of $\Phi(f)$. \\

We are now in position to give the Fock space representation of the algebra (3.1)-(3.4). As in the case of one scalar complex field two representations are possible: one on four component test functions and another one on two component test functions by the process of doubling the number of variables. We give here the second one. Let $f=\begin{pmatrix}f_1\\f_2\end{pmatrix} ;f_1,f_2 \in S(\mathbb R^4)$ and consider the scalar product given over the matrix $\underline\Delta^+$. Then

\begin{gather}
(A(f)\psi))(p_1,\ldots,p_n;q_1,\ldots,q_n)= \notag \\ =n^{-\frac{1}{2}}\sum_{i=1}^n
\begin{pmatrix}f(p_i)\\0 \end{pmatrix}\psi(p_1,\ldots,\hat p_i,\ldots,p_n;q_1,\ldots,q_m)+\\ \notag
+(m+1)^\frac{1}{2}(\begin{pmatrix}\bar f(-q)\\0\end{pmatrix},\psi(p_1,\ldots,p_n;q,q_1,\ldots,q_m))\\
(F^*(f))\psi)(p_1,\ldots,p_n;q_1,\ldots,q_n)= \notag \\ =n^{-\frac{1}{2}}\sum_{i=1}^n
\begin{pmatrix}0\\f(p_i)\end{pmatrix}\psi(p_1,\ldots,\hat p_i,\ldots,p_n;q_1,\ldots,q_m)+\\ \notag
+(m+1)^\frac{1}{2}(\begin{pmatrix}0\\ \bar f(-q)\end{pmatrix},\psi(p_1,\ldots,p_n;q,q_1,\ldots,q_m))
\end{gather}
and

\begin{gather}
(A^*(f)\psi))(p_1,\ldots,p_n;q_1,\ldots,q_n)= \notag \\ =m^{-\frac{1}{2}}\sum_{i=1}^n
\begin{pmatrix}\bar f(q_i)\\0\end{pmatrix}\psi(p_1,\ldots,p_n;q_1,\ldots,\hat q_i,\ldots,q_m)+\\ \notag
+(n+1)^\frac{1}{2}(\begin{pmatrix}\bar f(-p)\\0\end{pmatrix},\psi(p,p_1,\ldots,p_n;q_1,\ldots,q_m))\\
(F(f))\psi)(p_1,\ldots,p_n;q_1,\ldots,q_m)= \notag \\ =m^{-\frac{1}{2}}\sum_{i=1}^m
\begin{pmatrix}0\\f(q_i)\end{pmatrix}\psi(p_1,\ldots,p_n;q_1,\ldots,\hat q_i,\ldots,q_m)+\\ \notag
+(n+1)^\frac{1}{2}(\begin{pmatrix}0\\ \bar f(-p)\end{pmatrix},\psi(p,p_1,\ldots,p_n;q_1,\ldots,q_m))
\end{gather}
By linearity we can give a more compact Fock space representation for the scalar components written as $S(f)=\begin{pmatrix}A(f_1)\\F(f_2)\end{pmatrix},f=\begin{pmatrix}f_1\\f_2\end{pmatrix};S\begin{pmatrix}f\\0\end{pmatrix}=A(f),
S\begin{pmatrix}0\\f\end{pmatrix}=F(f)$. 
This representation as well as the causal field algebra for $S$ can be inferred from (3.11) and (3.14).\\
We encounter again (factorizable) zero vectors generated by $\underline\Delta^+$. For our case ($m=1$) they are given by $f_1+f_2=0$. As expected the zero vectors are generated by the equations of motion applied to the test functions. Alternatively we can start with test functions $ f=\begin{pmatrix}f_1 \\ f_2 \end{pmatrix} $ satisfying $ f_1+f_2=0 $ (as equation of motion) in which case there will be no zero vectors. A last remark: the reader should not be afraid of doubling the variables in $\psi$ from p to p and q; it is a very convenient way of bookkeeping for computations.      \\

\section{The Majorana field as component of the chiral multiplet}

The fermionic part of the supersymmetric chiral free field \cite{WB} consist of a complex Weyl multiplet $\chi_a(x), \quad a=1,2 $ with its formal (operator) adjoint $\bar\chi_{\bar a}=\chi_a^* $ assembled into the Majorana field $\chi=\begin{pmatrix}\chi_a \\ \bar\chi^{\bar a}\end{pmatrix},\quad \bar\chi^{\bar a}=\epsilon^{\bar a\bar b}\bar\chi_{\bar b}$. The causal Majorana free quantum field algebra is formally given by \cite{CSp}:

\begin{gather}
\{\chi_a(x),\chi_b(y)\}=im\epsilon_{ab}D(x-y) \\
\{\bar\chi^{\bar a}(x),\bar\chi^{\bar b}(y)\}=im\epsilon^{\bar a\bar b}D(x-y) \\
\{\chi_a(x),\bar\chi^{\bar b}(y)\}=\epsilon^{\bar b\bar c}\sigma_{a\bar c}\partial D(x-y) \\
\{\bar\chi^{\bar a}(x),\chi_b(y)\}=\epsilon_{bc}\bar\sigma^{\bar a c}\partial D(x-y)
\end{gather}
Here the notations are standard (the only deviation from for instance \cite{WB} is the different choice of the metric $(1,-1,-1,-1)$ instead of $(-1,1,1,1)$ together with $\sigma^0=\bar\sigma_0=1$ instead of $ \sigma^0=\bar \sigma_0=-1$). In particular the antisymmetric tensors $\epsilon_{ab}$ and $\epsilon^{ab};\quad a,b=1,2$ are defined through $\epsilon_{12}=\epsilon^{21}=-1$). Note that from the formal algebraic point of view (4.1)-(4.4) coincide with the propagator relations (9.11) in \cite{WB} if we replace the Pauli-Jordan commutator by the Feynman propagator.\\
It is not necessary to identify $\bar \chi _{\bar a} $ with the adjoint $\chi _a^* $. The identification is consistent with the equation of motion which is the Dirac equation for the fermion satisfying the Majorana condition (Majorana fermion) given below. A priori we will not assume them (see also the disscussion at the end of this section).\\
In order to give the Fock space representation of the (Weyl) spinor $\chi$ we find useful to use in parallel the companion (Majorana) bispinor notations (in fact we define the Weyl spinors through the Majorana's):        

\begin{equation}
\phi(x)=\begin{pmatrix}\phi_1(x)\\ \phi_2(x)\\ \phi_3(x)\\ \phi_4(x)\end{pmatrix}=\begin{pmatrix}\chi_1(x)\\ \chi_2(x)\\ \bar \chi^{\bar 1}(x)\\ \bar \chi^{\bar 2}(x)\end{pmatrix}=\begin{pmatrix}\chi_1(x)\\ \chi_2(x)\\ \chi_2^*(x)\\ -\chi_1^*(x) \end{pmatrix}
\end{equation}
where 

\begin{gather}
\phi_a(x)\equiv\chi_a(x),\quad a=1,2\\
\phi_b(x)\equiv (\chi^a)^*(x),\quad a=1,2, \quad b=4-a=3,4
\end{gather}
Here $(\chi^a)^*(x),a=1,2$ stays for $\bar\chi^{\bar a}(x),\bar a=\bar 1,\bar 2$. In the bispinor notation the formal (operator) Majorana condition is
\begin{equation}
\phi (x)=\mathcal{E}\phi^{*T}(x)
\end{equation}
where $T$ denotes transposition and $ \mathcal{E}=\begin{pmatrix}0 & \epsilon \\ -\epsilon & 0 \end{pmatrix}$. Here by $\epsilon$ we denoted the above mentioned antisymmetric tensor with lower indices. The transposition was introduced in order to preserve rules of matrix multiplication.\\
The relations (4.1)-(4.4) can now be written in an equivalent way:

\begin{equation}
\{\phi_a(x),\phi_b(y)\}=M_{ab}(-iD(x-y)),\quad a,b=1,\ldots,4
\end{equation}
with

\begin{equation}
M=\begin{pmatrix}-m\epsilon &i\sigma\partial\epsilon\\-i\bar\sigma\partial\epsilon &m\epsilon\end{pmatrix},\quad \epsilon=\begin{pmatrix}0 &-1\\1 &0\end{pmatrix}
\end{equation}
Remark that in (4.1)-(4.4) the indices $a,b$ run over one and two wheareas in (4.9) they run over one to four.\\
The Fourier transform of the matrix operator $M$ which we denote by $\tilde M$ is given by

\begin{equation}
\tilde M=\tilde M(p)=\begin{pmatrix}-m\epsilon & \sigma p\epsilon \\ -\bar \sigma p\epsilon & m\epsilon \end{pmatrix}
\end{equation}

Although it contains adjoints, the Majorana field describes a neutral field. This simplifies the matter of its Fock space representation. Indeed the process of doubling variables which we advocated for a complex field or multiplet components (see Section 3) is no longer necessary.\\

Let $f(x) \in S(\mathbb R^4)\otimes\mathbb C^4 $ be a (column) matrix of test functions in $S(\mathbb R^4)$ with Fourier transform $\tilde f(p)$. As in the preceding sections we omit the tilde from $\tilde f(p)$ writing just $f(p)$ for the Fourier transform. Let us introduce the notation

\begin{equation}
\phi(f)=\begin{pmatrix}\phi_1(f_1)\\ \phi_2(f_2) \\ \phi_3(f_3) \\ \phi_4(f_4)\end{pmatrix}
\end{equation}

Remark that as before in (4.12) the argument of $f$ on the left hand side is $x$ whereas on the right hand side the argument in $f_1,\ldots,f_4$ is $p$.\\
Now we can write down the representation of (4.9) and hence of (4.1) to (4.4) in the antisymmetric Fock space $\FF$ over $\FF^1 \supset E=S(\mathbb R^4)\otimes\mathbb C^4$. It is simply:

\begin{equation}
\begin{split}
&(\phi(f)\psi)(p_1,\ldots,p_n)=n^{-\frac{1}{2}} \sum_{i=1}^n(-1)^{i-1}f(p    _i)\psi(p_1,\ldots,\hat p_i,\ldots,p_n) \\ 
&+(n+1)^\frac{1}{2}\langle\bar f(-p),\psi(p,p_1,\ldots,p_n\rangle
\end{split}
\end{equation}
where $\langle\bar f(-p),\psi(p,p_1,\ldots,p_n)\rangle$ is defined up to some factors of $2\pi$ which we agree to consider included in $ \tilde M $ through
\begin{equation}
\langle f,g \rangle=(f,Mg)=\int\sum_{a,b=1}^4 \bar f_a(p)\tilde M_{ab}g_b(p)\Delta^+(p)dp
\end{equation}

A direct computation shows that the commutation relations (4.9)(and hence (4.1)-(4.4)) in their smeared form) are satisfied. We give a hint to the computation by checking it on the vacuum. Indeed let us identify the test function $f_a(x)$ with the bispinor $f_a(x)$ containing $f_a(x)$ as its only non vanishing component and write $\phi_a(f)=\phi(f_a),a=1,2,3,4 $. Then on the vacuum $\phi(f_a)$ retains only its creation part i.e. in Fourier space

\begin{equation}
\phi(f_a)\Omega=f_a(p_1)
\end{equation}
Now we distroy the vacuum and get

\begin{equation}
(\Omega,\phi(f_a)\phi(g_b)\Omega ) =\int f_a(-p)\tilde M_{ab}(p)g_b(p)\Delta^+(p)dp
\end{equation}
The same reasoning produces

\begin{equation}
(\Omega,\phi(g_b)\phi(f_a)\Omega ) =\int g_b(-p)\tilde M_{ba}(p)f_a(p)\Delta^+(p)dp
\end{equation}
where $(.,.)$ is the Fock space scalar product to be detected below. By taking here $p$ to $-p$ and adding (4.17) to (4.16) we obtain in this particular case (vacuum sandwich) the right anticommutation relation.  We use here the nice equalities $\bar \sigma \epsilon=\epsilon \sigma^T=-(\sigma \epsilon)^T $ to prove that $\tilde M^T(p)=-\tilde M(p) $. \\

Let us remark that although (4.13) looks very simple and apparently similar to the Fock space representation of the scalar neutral field it has some particularities which we discuss now. First (4.14) is not a scalar product in $S(\mathbb R^4)\otimes\mathbb C^4$ being NOT positive definite! This can be easily seen by remarking that the trace of $M$ vanishes. This seems at the first sight to be at odd with QFT-positivity. The answer follows from the peculiar structure of the Majorana fermion which although describing a neutral field doesn't satisfy $\phi^*(f)=\phi(\bar f)$ for $f \in S(\mathbb R^4)\otimes\mathbb C^4$. In order to verify that the QFT-positivity is satisfied we have to compute the two point function $\| \phi(f)\Omega \|^2=(\phi(f)\Omega,\phi(f)\Omega)$ which turns out to be positive definite. We give a hint leaving this computation to the interested reader. Indeed the explicit computation summarized below shows that the  true scalar product is not induced by the matrix $M$ but by another matrix $N$:

\begin{equation}
 N=\begin{pmatrix}i\bar \sigma \partial & -m \\ -m & i\sigma \partial \end{pmatrix}
\end{equation}
which in the Fourier variables reads

\begin{equation}
\tilde N=\tilde N(p)=\begin{pmatrix}\bar \sigma p & -m \\ -m & \sigma p \end{pmatrix}
\end{equation}
The point is that $\Delta^+(p)$ restricts $\tilde N= \tilde N(p) $ on the forward mass hyperboloid $p^2=m^2$ where $\tilde N $ is a $4\times 4$ positive definite matrix as we will prove below. But first let us show how we detect the matrix $N$. We write (4.9) in the form

\begin{equation}
\begin{pmatrix}\{\chi_a^*(x),\chi_b(y)\} §\quad \{\chi_a^*(x),(\chi^b)^*(y)\} \\ \{\chi^a(x),\chi_b(y)\} §\quad \{\chi^a(x),(\chi^b)^*(y)\} \end{pmatrix}=
M_{ab}(-iD(x-y))
\end{equation}
where the indices $a,b$ run over one and two on the l.h.s and over one to four on the r.h.s. of this equation. Now rearrange the matrix on the left hand side and verify that

\[ \{\phi_a(x),\phi_b(y) \}=
\begin{pmatrix}\{ \chi_a^*(x),\chi_b(y) \} § \quad \{\chi_a^*(x),(\chi^b)^*(y) \} \\ \{\chi^a(x),\chi_b(y) \} § \quad \{ \chi^a(x),(\chi^b)^*(y) \} \end{pmatrix}=N_{ab}(-D(x-y))
\]
with $a,b$ running over one to four as $\phi$ and one to two as $\chi $ indices and $N$ defined above. This produces the two point function in Fourier variables

\begin{equation}
(\phi_a(f)\Omega ,\phi_b(g)\Omega )=(\Omega ,\phi_a(f)^* \phi_b(g) \Omega )=
\int \bar f_a(p)\tilde N_{ab}(p)\Delta ^+(p)g_b(p)dp
\end{equation}
Introducing the scalar product

\begin{equation}
(f,g)=\int \bar f_a(p)\tilde N_{ab}(p)\Delta^+(p)g_b(p)dp
\end{equation}
it only remains to prove that the matrix $\tilde N $ when restricted to the forward mass hyperboloid is positive definite. In order to prove this result the reader can appeal to background knowledge on Dirac fermion (the method of Dirac projection operators) or use the following elementary result:\\

Consider the two by two positive definite matrix $A$ with unit determinant. Let 
\[
\mathbb A=\begin{pmatrix}A &cI \\cI &A^{-1}\end{pmatrix}
\]
where $I$ is the two by two unit matrix and $c=\pm 1$. Using the Hurwitz criterium it is easy to see that $\mathbb A$ is a four by four positive definite matrix. In fact the three and four determinants in the Hurwitz criterium vanish. Positive definiteness of $A$ induces positive definiteness of $\mathbb A $. \\

In order to apply this result to our situation concerning  $\tilde N$ we have only to scale it by the mass and use $(\bar \sigma p)(\sigma p)=p^2=m^2 $ as well as the positivity of $\sigma p$ on the forward mass-hyperboloid (det and trace of $ \sigma p $ are both positive!). \\

We will see later on that the Majorana positivity is strongy related to the Dirac fermion positivity. Using the result above it turns out that both are consequences of Weyl positivity, i.e. the positivity of $\sigma p$ considered on the forward mass hyperboloid. But before relating the Majorana to the Dirac fermion let us remark that for the scalar product under consideration on the space of test functions $S(\mathbb R^4)\otimes\mathbb C^4 $ there are zero vectors of the form $f=\begin{pmatrix}f_a  \\ \bar f^{\bar a}\end{pmatrix}$ satisfying the Majorana equations of motion (which is the Dirac equation for the Majorana fermion). They are harmless and can be easily factorized. Certainly it is possible to start the (antisymmetric) Fock space construction from the beginning with the test functions $f$ in $ S(\mathbb R^4)\otimes \mathbb C^4 $ which are at the same time solutions of the Majorana equations equipted with the scalar product $(.,.)$ in which case there will be no zero vectors. Remark the perfect agreement between the two types of component fields of the supersymmetric chiral multiplet stemming from the fact that both multiplet components: the scalar one consisting of the fields $A,F$ and $A^*,F^*$ as well as Majorana consisting of the Weyl spinors $\chi $ and $ \bar \chi $ are defined on test functions which may or may not satisfy equations of motion.\\

 By defining $\phi^*(f) $ in a similar way as in (4.9) (which is in fact a reorganization of it) we can realize the rigorous operatorial equality $ \phi (f)^*= \phi^*(\bar f) $ together with the smeared version of (4.8) where $ \phi (f)^* $ is the operator adjoint of $ \phi (f) $. A Fock space representation of $ \chi(f) $ can be read off from (4.13) by setting $f=\begin{pmatrix}f_a  \\ 0 \end{pmatrix}$. It is a representation on four component test functions because the quantum field $ \chi (f) $ was introduced over the quantized Majorana field $ \phi (f) $. A representation over two component test functions is also possible and would correspond to the canonical quantization of the classical Weyl field (see for instance \cite{CSp}). In this representation we can verify the operator identity $\chi(f)^*= \chi^*(\bar f) $ .\\

In order to appreciate the Majorana fermion Fock space representation (4.13) let us put aside the Dirac fermion $\phi^D(f)$ Fock space representation. There are at least three ways to obtain useful Fock space representations for Dirac fermions, the most common \cite{SW} being the representation on n-point antisymmetric functions of the form $\psi(x_1,e_1;\ldots;x_n,e_n), \, e_i=\pm 1, \, i=1,\ldots,n $. A more convenient representation is obtained by the process of doubling the number of variables as this was already done for the complex scalar field in section 3 (we use here the nice exposition in \cite{B} which was influential for our paper; see also \cite{K}). It is:

\begin{equation}
\begin{split}
&(\phi^D(f)\psi)(p_1,\ldots,p_n;q_1,\ldots,q_m)= \\
&=n^{-\frac{1}{2}}\sum_{i=1}^{i=n}(-1)^i f(p_i)\psi(p_1,\ldots,\hat p_i,\ldots,p_n;q_1,\ldots,q_n)+ \\
&+(m+1)^\frac{1}{2}(\bar f(-q),\psi(p_1,\ldots,p_n;q,q_1,\ldots,q_m))_t \\
&(\phi^{D*}(f)\psi)(p_1,\ldots,p_n;q_1,\ldots,q_m)= \\
&=m^{-\frac{1}{2}}\sum_{i=1}^{i=n}(-1)^i f(q_i)\psi(p_1,\ldots,p_n;q_1,\ldots,\hat q_i,\ldots,q_n)+ \\
&+(n+1)^\frac{1}{2}(\bar f(-p),\psi(p,p_1,\ldots,p_n;q_1,\ldots,q_m))
\end{split}
\end{equation}
where the scalar product is given by the same matrix $N$ as above. 
Remark the transpose $t$ of the scalar product in the first relation. It plays an important role as compared with the Majorana representation (4.13). Indeed in the process of veryfing the field anticommutator relation in the Dirac case it is this transposition that forces $\Delta^+ (p)$ into $\Delta^- (p)$ producing finally $\Delta(p)$ and as such causality of the Dirac field. In comparisson to this the Majorana causality is generated by (4.13) as a consequence of the relations  $\bar \sigma \epsilon=\epsilon \sigma^T=-(\sigma \epsilon)^T $ . The (neutral) Majorana fermion is by its reality condition more rigid than the (charged) Dirac fermion.\\

Let us remark in passing that yet another Fock space representation of the Dirac fermion is possible; instead of doubling the number of variables we can split the test functions in the positive and negative regions of the time variable. The resulting Fock space representation is simpler than that in (4.23). We do not need it for the purposes of this paper but recommend it as an exercise being the simplest Fock space representation of the Dirac fermion one can ever think about. The interested reader can find details in \cite{CScf} where the case of one dimensional chiral fermions was worked out explicitely or in \cite{S} where it is applied for concrete computations of scattering processes\\

Before ending this section on Majorana fields let us first remark that our definition (4.1) to (4.4) of the causal Majorana algebra can be recovered as in \cite{CSp} by canonical quantization. It is now interesting to point out that yet another definition is possible. Indeed we can build up the Majorana fermion from the Dirac fermion at the CLASSICAL level as usually done in the literature on the subject (see for instance \cite{W}) and quantize it as Dirac fermion! The resulting multiplet satisfies again causality and positivity and the interested reader can easily write down the Fock space representation of it following the methods presented above.  

\section{Applications, Conclusions and Remarks}

We have given Fock space realizations of the scalar and Majorana components of the chiral superfield $ \Phi(x,\theta,\bar\theta)$. We can now look at the chiral superfield as being represented in the tensor product space of these two Fock spaces. It would be interesting to find a Fock space realization of the chiral superfield over (test) functions depending on commutative and noncommutative variables with definite symmetry properties. Although our efforts in this paper do not provide results of this type, they are sufficient for a supersymmetry extension of the causal perturbative method in quantum field theory. This method was invented by Epstein and Glaser by
the end of sixties \cite{EG} and was subsequently extended in several directions (for a recent report see \cite{Sg}). Without going into details we show here that the non-renormalization theorem for the chiral model (the
Wess-Zumino model) can be easily derived by the supersymmetric generalization of the above mentioned renormalization method. Indeed, working in the frame of our Fock space realization of the causal multiplet components by looking at $\theta$ and $\bar\theta$ as bookkeeping variables we realize that all formal computations which are carried out in supersymmetry concerning the Wess-Zumino model can be carried out in a rigorous way too in our Fock-space approach without any reference to the functional integral and formal time-ordered products. What is needed in the frame of the causal perturbation method are only Wick products, Wick powers and the Wick theorem (in its slightly extended form known as the ``zero theorem" \cite{EG}). All this constructions are easily achieved in our framework. For convenience we use here a simplified Epstein-Glaser method which has the advantage of reducind the distribution theoretic aspects to an extention problem directly related to Wick products (this variant is nicely described in \cite{Pr}). The interesting point is now that mass and coupling constant renormalization are solely generated by Wick products of either $\Phi(x,\theta,\bar\theta)$ or $\bar\Phi(x,\theta,\bar\theta)$ factors in the perturbation expansion. No mixed terms in $ \Phi $ and $ \bar\Phi $ do appear. By the straightforward supersymmetric extension of the zero theorem it follows that in any order of perturbation theory only TREE contractions survive. Indeed all closed-loop diagramms vanish when they contain superfields of the same chirality. This follows as usual from the fact that $\delta^2(\theta-\theta ')=\delta(\theta -\theta ')\delta(\theta -\theta ')=\delta(0)=0$ (see \cite{WB} p.64). Altogether this implies the absence of mass and coupling constant renormalization. The argument is of deceptive simplicity. \\

Concerning the discussion above in which we apparently repeated old arguments the reader might ask himself what is the new point of our approach. To answer this question we stress the fact that our approach is genuine a quantum one  having no a priori connection to the classical case. There are no classical supersymmetric fields and no functional integration is used to generate time-ordered products, Feynman diagramms etc. But it has to be mentioned that the formal abgebraic structure of the theory is not changed. In particular two-point functions computed in our Fock-space setting are formally consistent with the corresponding propagators \cite{WB}. These remarks concern the older component approach but what was more important for us they  equally apply to the superspace approach segregating  vanishing contributions to the mass and coupling constant renormalization. We close this paper by remarking that present considerations can be extended to other models like vector superfields or even $N=1$ supersymmetric gauge theories (for some details see \cite{CSg}).\\
Acknowledgement \\
One of the authors (F.C.) thanks Matthias Schork for discussions.\\


\begin{thebibliography}{99}
\bibitem {WB} J. Wess, J.Bager, Supersymmetry and Supergravity, 2nd edition, Princeton University Press, 1992
\bibitem {W} S. Weinberg, The Quantum Theory of Fields, vol.III, Cambridge University Press, 2000
\bibitem {CSp} F. Constantinescu, G. Scharf, Causal approach to supersymmetry: chiral superfields, preprint Z\"urich, 2000
\bibitem {EG} H. Epstein, V. Glaser, Annales Inst. Poincaré A 19 (1973) 211
\bibitem {S} G. Scharf, Finite Quantum Electrodynamics, 2nd edition, Springer Verlag, New York, Heilelberg, Berlin, 1995
\bibitem {Sg} G. Scharf, Quantum Gauge Theories - a True Ghost Story, Wiley, New York, 2001
\bibitem {West} P. West, Introduction to Supersymmetry and Supergravity, Extended second edition, World Scientific, 1990
\bibitem {J} R. Jost, The General Theory of Quantized Fields, American Mathematical Society, 1965
\bibitem {SW} R.F. Streater, A.S. Wightman, PCT, Spin and Statistics and All That, Benjamin, 1964
\bibitem {BLT} N.N. Bogoliubov, A.A. Logunov, I.T. Todorov, Mathematical Methods in Quantum Field Theory, Dover Publications, 1975
\bibitem {B} H-J. Borchers, Translation Group and Particle Representations in Quantum Field Theory, Springer, 1996
\bibitem {K} D. Kazhdan, Introduction to QFT in "Quantum Fields and Strings: A course for matematicians", vol 1, American Mathematical Society, 1999
\bibitem {CScf} F. Constantinescu, G. Scharf, Lett. Math. Phys. 52
(2000) 113
\bibitem {Pr} D. Prange, Kausale St\"orungstheorie und differentielle Renormierung, Diplomarbeit, II. Institut f\"ur Theoretische Physik, Universit\"at Hamburg, 1997
\bibitem {CSg} F. Constantinescu, G. Scharf, Quantized hermitean superfields, Preprint Z\"urich, 2001
\end{thebibliography}
\end{document}